\documentclass[prr,aps,twocolumn,%
showpacs,preprintnumbers,amsmath,amssymb]{revtex4-1}
\usepackage[utf8]{inputenc}
\usepackage{graphicx}
\usepackage[breaklinks, colorlinks=true]{hyperref}
\usepackage{bm}
\usepackage{slashed}

\usepackage{physics}

\newcommand{\calH}{\mathcal{H}}
\newcommand{\calM}{\mathcal{M}}
\newcommand{\bepsilon}{{\bm \epsilon}}
\newcommand{\bnabla}{{\bm \nabla}}
\newcommand{\be}{{\bm e}}
\newcommand{\bj}{{\bm j}}
\newcommand{\bk}{{\bm k}}
\newcommand{\br}{{\bm r}}
\newcommand{\bq}{{\bm q}}
\newcommand{\bs}{{\bm s}}
\newcommand{\bA}{{\bm A}}
\newcommand{\bB}{{\bm B}}
\newcommand{\bC}{{\bm C}}
\newcommand{\bE}{{\bm E}}
\newcommand{\bJ}{{\bm J}}
\newcommand{\bS}{{\bm S}}

\begin{document}
\title{Implications of the helicity conservation for on-shell and
  off-shell chirality production}

\author{Kenji Fukushima}
\email{fuku@nt.phys.s.u-tokyo.ac.jp}
\affiliation{Department of Physics, The University of Tokyo, 
  7-3-1 Hongo, Bunkyo-ku, Tokyo 113-0033, Japan}

\author{Chengpeng Yu}
\email{yu.chengpeng@nt.phys.s.u-tokyo.ac.jp}
\affiliation{Department of Physics, The University of Tokyo, 
  7-3-1 Hongo, Bunkyo-ku, Tokyo 113-0033, Japan}

\begin{abstract}
  We analyze the helicity conservation
  following from the axial Ward identity for the massless fermion.
  We discuss the pair production of a fermion,
  $\ell$, and an anti-fermion, $\bar{\ell}$, from the annihilation
  process of photons, $\gamma$'s and conclude that the chirality production is prohibited in any annihilation
  process of on-shell $\gamma$'s.  We demonstrate this selection rule in
  the tree-level process of $\gamma+\gamma\to \ell+\bar{\ell}$.  We
  next consider the off-shell effect on the chirality
  production in the presence of the background magnetic field.  We
  point out that the particle conversion process under the
  magnetic field induces the polarization of an incident $\gamma$, which can also be interpreted intuitively from the helicity conservation.
\end{abstract}
\maketitle

\section{Introduction}

Chiral anomaly has vital impacts on hadron physics founded on quantum
chromodynamics (QCD), but it is still a challenge how to identify experimental observables to probe chiral
anomaly.  In QCD chiral anomaly is
driven by topologically winding configuration called the instanton,
but the instanton is rather a mathematical device only to provide a
Euclidean interpretation of quantum tunneling.  Thus, there is no way
to detect the instanton as a physical excitation in the laboratory.
To circumvent the difficulty, we can investigate special environments under a strong magnetic field in which chiral anomaly could be macroscopically manifested.  As a
matter of fact, the magnetic field as a device to examine intriguing academic exercises has been
discussed repeatedly and one well-known example is found in the
formulation of the chiral anomaly matching between the infrared and
the ultraviolet sectors; see Ref.~\cite{Smilga:2000ek} for a review
including early-days discussion.  We further emphasize the experimental relevance of
strong magnetic field beyond the academic exercise.  (Readers
interested in the academic exercise can consult
Ref.~\cite{Fukushima:2018grm} for a pedagogical review.)  In the
relativistic heavy-ion collision experiment, hot and dense matter
should be exposed to a gigantic magnetic field whose energy scale is 
comparable to that of the strong interaction, that is, $\sqrt{eB}\sim \Lambda_{\text{QCD}}$.  In central cores of
neutron stars, moreover, the magnetic flux is squeezed and a sizable
effect on the ground state properties of quark matter is expected~\cite{Fukushima:2007fc,Noronha:2007wg,Ferrer:2012wa}.
See a review~\cite{Miransky:2015ava} for theoretical developments and
application in nuclear, astro-nuclear, and condensed matter physics.
The comprehensive summary of the state-of-the-art status is found in a
recent review~\cite{Hattori:2023egw}.

The advantage in dealing with extreme matter with strong magnetic
field is that the chiral anomaly can give rise to macroscopic effects through the anomalous transport; see Ref.~\cite{Kharzeev:2015znc} for a review
about the chiral transport effects particularly in the
heavy-ion collision.  One of the intuitively understandable ways to formulate
the anomaly-induced transport is based on the helicity conservation derived from the axial Ward identity.  More specifically, the
magnetic helicity in the gauge sector is converted into chirality in
the matter sector, and vice versa.  This formulation can be generalized
to systems including the fluid vorticity~\cite{Avdoshkin:2014gpa}.
For the classical description of the helicity conservation for
electromagnetic and chiral fluids, see a recent work~\cite{Manuel:2022tck}.

We emphasize that the argument of the helicity conservation has a
close connection to optical physics.  It is often the case that the
magnetic helicity is assumed to be conserved in optics.
In fact, the symmetry-based argument for the helicity conservation in
free Maxwell equations is dated back to the classic paper by Calkin in
1965~\cite{10.1119/1.1971089}.  The famous
``zilch''~\cite{10.1063/1.1704165} can also be derived as a Noether
current.  In recent discussions (see Ref.~\cite{2012NJPh...14l3019C}
for example) it is common to utilize a field-theoretical reformulation
of the helicity conservation law in terms of not only the vector
potential $\bA$ but also the dual vector potential $\bC$.  Although
the conserved nature of the optical helicity can be confirmed explicitly in this
reformulation, such a theory with dual fields is not necessarily
equivalent with the original theory.  If the continuity equation is
written for the magnetic helicity in terms of $\bA$, a residual term,
$\bE\cdot\bB$, appears and this unwanted (but essential for the chiral
anomaly) term is cancelled by the dual contribution in terms of $\bC$ (see the next session for technical details).

In the quantum level, the helicity conservation should be considered
from a solid foundation of quantum field theory, that is, the axial
Ward identity.  Usually, the axial Ward identity is regarded as a
relation for explicit symmetry breaking due to the chiral anomaly, and
nevertheless, it can be taken for the conservation law.  Indeed, it is
an interesting subject to define the symmetry generator corresponding
to the helicity conservation in electromagnetic dynamics, and the
non-invertible symmetry has been recognized to make the symmetry
generator invariant under large gauge
transformation~\cite{Harlow:2018tng,Yamamoto:2023uzq}.  In this work
we will not consider quantized magnetic flux, so that we can focus on
the ordinary magnetic helicity only.  Then, an intriguing question is what microscopic process is responsible for transforming the magnetic
helicity into the matter helicity (or the chirality).  We already know
that the parallel configuration of $\bE$ and $\bB$ allows for the
Schwinger mechanism with a fermion, $\ell$, and an anti-fermion,
$\bar{\ell}$, in such a way that the net chirality of $\ell$ and
$\bar{\ell}$ is $\pm 2$ as closely discussed in
Ref.~\cite{Copinger:2018ftr,Taya:2020bcd}.  It is generally an instructive exercise to
think of the Schwinger mechanism with $\bE\parallel\bB$ (see Ref~\cite{Dunne:2004nc} for a review), but the
problem is that the experimentally accessible $\bE$ is not yet strong enough to trigger the Schwinger
effect in the vacuum (see Ref.~\cite{tajima2002zettawatt} for the history of laser-intensity developments).  In condensed matter physics, the electron transport provides us with a feasible experimental
setup to detect the chiral anomaly in media.  As
pointed out in Refs.~\cite{Kharzeev:2007jp,Fukushima:2008xe}, a finite
chirality imbalance under $\bB$ should induce a nonzero electric
current, $\bj\propto\bB$.  Then, this idea is applied to the
condensed matter system and the negative magnetoresistance was
found to be a signature for the chiral anomaly~\cite{Son:2012bg}.  In
a microscopic view, to measure the electric conductivity $\sigma$, a
finite $\bE$ is imposed in a medium and $\bB\parallel\bE$ is
introduced externally, which produces the chirality through the
Schwinger mechanism.  The produced chirality gives rise to
an electric current along $\bB$, and $\sigma$ is expected to have an
anomalous component that increases with increasing $|\bB|$ (i.e., the
positive magnetoconductivity or the negative magnetoresistance).  This
theoretical expectation has been confirmed by experiments (among which
the first one was reported in Ref.~\cite{Li:2014bha}) as well as the
QCD calculations, see Ref.~\cite{Astrakhantsev:2019zkr} for the
first-principles numerical simulation and
Refs.~\cite{Fukushima:2017lvb,Fukushima:2019ugr} for the resummed
perturbation theory.  These are all intriguing progresses, but the
theoretical calculations are highly complicated and the physical
interpretation is not really transparent.  There are several theoretical
examples~\cite{Landsteiner:2014vua,Jimenez-Alba:2015awa,Sun:2016gpy,Fukushima:2021got,Sogabe:2021wqk}
which imply that the negative or positive magnetoresistance may have
some other origins apart from the chiral anomaly.

This motivates us to seek for an alternative channel to detect the chiral
anomaly.  The Schwinger mechanism is a nonperturbative process,
and the theoretical analysis is not straightforward.  So, a simpler perturbative process for the chirality production, if any, would be more convenient for theoretical approaches.  Then, one immediate example would be a perturbative treatment of the particle production from photons whose energies are sufficiently large.  If a finite chirality is produced in
a simple scattering process like $\gamma+\gamma\to\ell+\bar{\ell}$, it would be
much easier to control the energetic photon beam than handling the constant
strong electric field.  This is the question that we would address in the first part of this work.  We next proceed to the mixed case with both on-shell and off-shell $\gamma$'s, for instance,
$\gamma \to \ell(B)+\bar{\ell}(B)$, in the magnetic environment.
Here, $\ell(B)$ and $\bar{\ell}(B)$ represent the fermion and the anti-fermion with multiple scatterings with off-shell photons from background $\bB$.
Technically speaking, we should utilize the fermion propagator under $\bB$ so that we can effectively resum infinite diagrams for the off-shell photon scattering.  In the former case with on-shell $\gamma$'s only and also in the latter case with off-shell $\gamma^\ast$'s, we find interesting nontrivial results and
we emphasize that the helicity conservation gives us a useful insight to deepen our understanding about these results.

This paper is organized as follows.
In Sec.~\ref{sec:helicity} we make a brief overview about the helicity conservation.  We first introduce conventional discussions in optics and then turn to the axial Ward identity and the chiral anomaly.
Section~\ref{sec:on-shell} is devoted to our main discussion about the possibility of the chirality production from the on-shell photon annihilation process.
We develop a general analysis based on the helicity conservation, and then confirm our conclusion with the diagrammatic calculation.
In Sec.~\ref{sec:polarizer} we focus on a slightly different setup with external magnetic field.
We apply the helicity conservation argument for the pair production process from the energetic photon in the magnetic field.
The production rate depends on the polarization of the injected photon, which implies that the external magnetic field may well be a polarizer.
Finally we conclude this work in Sec.~\ref{sec:conclusions}.

\section{Helicity Conservation}
\label{sec:helicity}

We will make a brief overview of the helicity conservation in two contexts of optics and high-energy physics.  First, let us see what is called the ``optical helicity'' in the field of quantum
optics.  The
most natural quantity to be identified as the helicity is the magnetic
helicity originally known in plasma physics defined by
\begin{equation}
  H_M = \int d^3\br\, \calH_M = \int d^3\br\, \bA\cdot\bB\,.
\end{equation}
Although the integrand, $\calH_M$, is not gauge invariant, the
integrated quantity, $H_M$, is gauge invariant up to the surface
contribution.
This quantity is supposed to be a conserved charge, and we should consider the spatial components for the conserved current, namely,
$\bs=\bE\times\bA+A_0\bB$.  However, the current is not strictly conserved, i.e., the Maxwell equations lead to
\begin{equation}
  \frac{\partial\calH_M}{\partial t} + \bnabla\cdot\bs =
  2\bE\cdot\bB\,.
  \label{eq:magnetic_continuity}
\end{equation}
This relation implies that the magnetic helicity could be regarded as a
conserve charge as long as $\int d^3\br\,\bE\cdot\bB=0$, which is the case for
common electromagnetic waves as solutions of the free Maxwell equations.

It has been
proposed to generalize the above expression including the dual vector
potential $\bC$ so that the continuity equation can be exact in the
absence of matter~\cite{2012NJPh...14l3019C,Crimin_2019}.
Here, $\bC$ is defined in a dual manner with $\bB$ replaced by $\bE$.
In general cases with $\bnabla\cdot\bE\neq 0$, however, we need to perform the Helmholtz decomposition, i.e.,
$\bE=\bE_{\perp}+\bE_{\parallel}$.
Then, we see $\bnabla\cdot\bE_{\perp}=0$, so that we can define $\bC$ to satisfy $\bE_{\perp}=-\bnabla\times\bC$.
Using $\bC$, we can express the generalized optical helicity as
$\calH_{\text{opt}} = \frac{1}{2}(\bA\cdot\bB - \bC\cdot \bE_{\perp})$
and the current part as
$\bs_{\text{opt}}=\frac{1}{2}(\bE_{\perp}\times\bA
+\bB\times\bC)$ in the $A_0=0$ gauge.
With some calculations, we can prove that $\bE\cdot\bB$ in the right-hand side of
Eq.~\eqref{eq:magnetic_continuity} is cancelled by the additional contribution from the dual counterpart.  The advantage of such a formulation is not only the
exact continuity equation but also the manifest duality between $\bB$
and $\bE$.  That is, this generalized optical helicity takes account
of not only the magnetic helicity but also the electric helicity.
Interestingly, it is also known that this dual formulation can
incorporate the violation of the helicity conservation in the presence of matter, which is expressed as~\cite{PhysRevA.93.023840}
\begin{equation}
  \frac{\partial\calH_{\text{opt}}}{\partial t}
  +\bnabla\cdot\bs_{\text{opt}}
  = \frac{1}{2} ({\bm g}\cdot\bnabla\times \bC
  +\bC\cdot\bnabla\times{\bm g})\,,
  \label{eq:opt_cont}
\end{equation}
where ${\bm g}$ is defined from the matter current, $\bj$, that is,
$\bj_{\perp}=\bnabla\times{\bm g}$.

This non-conservation relation in the presence of matter has a suggestive
form, but it should be noted that the cancellation of the right-hand
side of Eq.~\eqref{eq:magnetic_continuity} implies the absence of the
anomalous effects by construction of the dual formulation.  Since we are
interested in the helicity transmutation between matter and gauge fields through the chiral
anomaly, the conventional magnetic helicity is a more suitable
quantity for our purpose.

It is time to turn our attention to high-energy physics.  If the
fermion mass is sufficiently small, we can give a more
quantum-field-theoretical foundation to
Eq.~\eqref{eq:magnetic_continuity}.  The axial Ward identity is an
exact relation associated with the anomalous violation of
$U(1)_{\mathrm{A}}$ symmetry.  The would-be conserved current, i.e.,
the axial current should satisfy the following relation:
\begin{equation}
  \partial_\mu j_5^\mu = -\frac{e^2}{16\pi^2}
  \epsilon^{\mu\nu\alpha\beta} F_{\mu\nu}F_{\alpha\beta} +
  2m\bar{\psi}i\gamma_5\psi\,.
\end{equation}
Let us assume that we can drop the last term proportional to the mass
$m$ if the system has only a massless fermion.  It would be instructive
to use a more familiar notation for the electromagnetic fields, $\bE$,
and $\bB$, instead of the field strength tensor, $F_{\mu\nu}$.  Then,
we can rewrite the above identity as
\begin{equation}
  \frac{d N_5}{d t} + \int_{\partial V} d\bS\cdot \bj_5
  -\frac{e^2}{2\pi^2}\int_V d^3\br \,\bE\cdot\bB = 0\,,
  \label{eq:integ_awi}
\end{equation}
after taking the integration over the spatial region $V$.  In general,
the last term is not necessarily an integer.  If spacetime is a direct
sum of $T^2\otimes T^2$ in the electric and the magnetic
sectors, the boundary condition quantizes the field flux in a form of
the first Chern number, that is, if the electric field is along the
$z$ direction, the integrated quantity is
\begin{equation}
  c_1^E = \frac{e}{2\pi}\int_{T^2} dx^0 dx^z\, F_{0z} \in \mathbb{Z}\,.
\end{equation}
In the same way the magnetic sector leads to $c_1^B\in\mathbb{Z}$.
According to the standard argument in the chiral anomaly, we assume
that the surface term can be discarded, and the time integration of
the axial Ward identity gives
\begin{equation}
  \Delta N_5 = N_5(t=\infty) - N_5(t=-\infty) = 2Q_{\mathrm{W}}
\end{equation}
with the topological winding number
$Q_{\mathrm{W}}=c_1^E c_1^B \in\mathbb{Z}$.  We can readily establish
an intuitive interpretation for the above relation.  Every time the
left-handed particle is converted into the right-handed particle via
the spectral flow [or equivalently, a right-handed particle and a
right-handed anti-particle (an anti-particle of the left-handed
particle) is created], the chirality changes by the unit of two.
In the next section we will encounter an example in which $H_M$ is integer (which is a similar situation as argued in Ref.~\cite{Hirono:2016jps}) and thus $Q_W$ is not, and then the physical interpretation becomes nontrivial.

\section{On-shell Photon Processes}
\label{sec:on-shell}

We shall look into the perturbative regime in which $Q_W\notin\mathbb{Z}$.  We first
discuss the general analysis using the helicity conservation, and next
go into a simple example of the diagrammatic calculation.

\subsection{General analysis}
For our discussions, it is convenient to transform the axial Ward
identity in the following form:
\begin{equation}
  \frac{d}{dt} \biggl( N_5 + \frac{e^2}{4\pi^2} H_M \biggr)
  + \int_{\partial V} \bigl( \bj_5 + \bj_{\mathrm{EM}} \bigr)\cdot d\bS = 0\,,
  \label{eq:axialWI}
\end{equation}
where
\begin{equation}
  \bj_{\mathrm{EM}} = \frac{e^2}{2\pi^2} A_0\bB
  + \frac{e^2}{2\pi^2} \bE\times \bA \,.
\end{equation}
The first term is nothing but the chiral magnetic effect if $A_0$ is identified as the chemical potential.
The quantization of the gauge field is crucial for following discussions in
the perturbative regime.  We adopt the Coulomb gauge with $A_0=0$, and
the physical transverse fields are expanded in terms of the annihilation/creation operators as
\begin{equation}
  \bA = \sum_{\sigma=\pm}\int\frac{d\bk}{(2\pi)^3}
  \frac{1}{\sqrt{2\omega}} \Bigl[ \bepsilon_\sigma(\bk)\,
  \hat{a}_{\sigma,\bk}\, e^{-i\omega t+i\bk\cdot\br}
  + \text{h.c.} \Bigr]\,.
\end{equation}
For the on-shell photon $\omega=|\bk|$ represents the photon energy.
We will see the explicit form of the polarization vectors, $\bepsilon_\sigma(\bk)$, in later discussions.
Then, we can immediately find an expression for the quantized magnetic
helicity, that is,
\begin{equation}
  H_M = \sum_{\sigma,\sigma'} \int \frac{d\bk}{(2\pi)^3}
  (-i\hat{\bk}) \cdot \bepsilon_\sigma^\ast(\bk) \times  
  \bepsilon_{\sigma'}(\bk)\, \hat{a}_{\sigma,\bk}^\dag  
  \hat{a}_{\sigma',\bk}\,.  
\end{equation}
Similar calculations give us a quantized form of the current,
$\bJ_{\text{EM}}=\int_V \bj_{\text{EM}}\,d\br$, as follows:
\begin{equation}
  \bJ_{\text{EM}} = \sum_{\sigma,\sigma'} \int \frac{d\bk}{(2\pi)^3}
  (-i) \bepsilon_\sigma^\ast(\bk) \times 
  \bepsilon_{\sigma'}(\bk)\, \hat{a}_{\sigma,\bk}^\dag 
  \hat{a}_{\sigma',\bk}\,. 
\end{equation}
The vector $\bk$ indicates the momentum carried by quantized photon, or
$\hat{\bk}$ represents the direction of photon propagation.  Then,
without loss of generality, two physical polarizations are chosen as
transverse vectors corresponding to the right/left circular polarizations (denoted by $\sigma=\pm$, respectively) orthogonal to
$\hat{\bk}$, and the magnetic helicity and the flux simplify as
\begin{align}
  H_M &= \int \frac{d\bk}{(2\pi)^3}\, 
  \bigl( \hat{a}_{+,\bk}^\dag \hat{a}_{+,\bk}
  - \hat{a}_{-,\bk}^\dag \hat{a}_{-,\bk} \bigr)\,,\\
  \bJ_{\text{EM}} &= \int \frac{d\bk}{(2\pi)^3}\, \hat{\bk}
  \bigl( \hat{a}_{+,\bk}^\dag \hat{a}_{+,\bk}
  - \hat{a}_{-,\bk}^\dag \hat{a}_{-,\bk} \bigr)\,. 
\end{align}
This implies that $\langle\text{phys}| H_M |\text{phys}\rangle
= N_+ - N_- \in \mathbb{Z}$ where $|\text{phys}\rangle$ is a state
with $N_\pm$ on-shell photons with $\sigma=\pm$.

\begin{figure}
  \includegraphics[width=0.45\columnwidth]{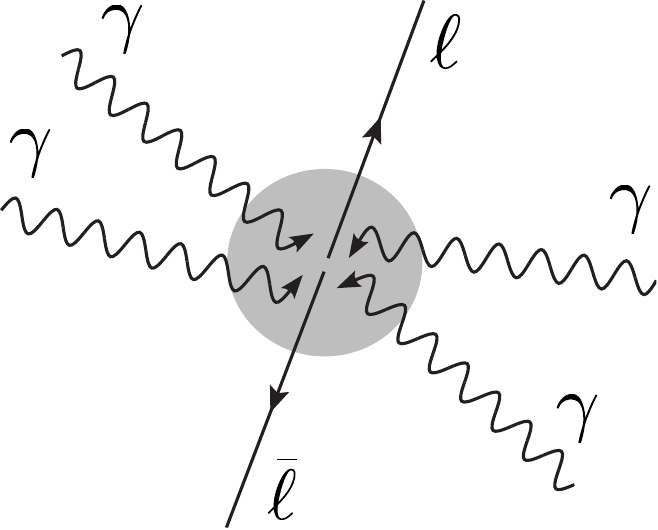}
  \caption{Schematic figure of annihilation processes from arbitrary
    number of photons to pairs of fermions and anti-fermions.  In this
    type of processes only the channels with vanishing helicity are
    possible.}
  \label{fig:scattering}
\end{figure}

We are ready to deduce the implication from the axial Ward
identity~\eqref{eq:axialWI}.  It is obvious that the magnetic helicity
cannot be converted into the matter chirality.  Both $N_5$ and $H_M$
are integral quantized, while a coefficient $e^2/(4\pi^2)$, which is not necessarily a rational number,
remains in Eq.~\eqref{eq:axialWI}.  This is a crucial difference from
the argument of the Chern number for which the quantization rule includes the coefficient.  Thus, it is impossible to increment $N_5$
in response to the change of $H_M$, which is prohibited by non-integer $e^2/(4\pi^2)$.
Interestingly, we can also make sure that $H_M$ is unable to change by
injecting $\bj_{\text{EM}}$.  To see this, let us consider a concrete
example, $\gamma+\gamma\to \ell+\bar{\ell}$.  If both incident
$\gamma$'s are polarized along the momentum direction, $H_M$ would
increase by the unit of two when two $\gamma$'s pass into the integration
region $V$ on the one hand.  When this happens, on the other hand, the
total $\bj_{\text{EM}}$ integrated over $\partial V$ is vanishing.
This argument clearly shows that $\bj_{\text{EM}}$ cannot compensate
for non-integer $e^2/(4\pi^2)$ in front of $H_M$ in the annihilation
process.  Therefore, we conclude that the annihilation process is
possible only in the channel with $\Delta N_5=\Delta H_M=0$, i.e., there is no
chirality production.  In general, for any perturbative
processes in which arbitrary number of $\gamma$'s annihilate into
pairs of fermions and anti-fermions as sketched in
Fig.~\ref{fig:scattering}, the magnetic helicity before the
annihilation is $(e^2/4\pi^2)(N_+-N_-) \notin\mathbb{Z}$, and the
chirality after the pair production is $N_5\in\mathbb{Z}$, and they
can never be equated unless $N_+-N_-=N_5=0$.  This generalization is a
statement of our no-go theorem stating that a nonzero chirality cannot be
produced with any on-shell photon annihilation.

\subsection{Lowest-order diagram}

Let us check our no-go theorem in a simple example, i.e., the
tree-level amplitude for
$\gamma(\bk_1)+\gamma(\bk_2)\to\ell(\bq_1)+\bar{\ell}(\bq_2)$.  We
denote the incident photon helicity polarization by $\sigma_1$ and $\sigma_2$
and the outgoing fermion spin polarization by $s_1$ and $s_2$.  The
scattering amplitude is then expressed as
$\calM_{\sigma_1 \sigma_2}^{s_1 s_2}=[\bepsilon_{\sigma_1}(\bk_1)]^\mu
[\bepsilon_{\sigma_2}(\bk_2)]^\nu \calM_{\mu\nu}^{s_1 s_2}$.  The
simple computation results in the following amplitude,
\begin{equation}
  \begin{split}
    & i\calM_{\mu\nu}^{s_1 s_2} = -ie^2 \bar{u}^{s_1}(q_1) \\
    &\times \biggl[
    \frac{\gamma_\mu (\slashed{q}_1 - \slashed{k}_2)\gamma_\nu}
    {(q_1-k_2)^2 + i\eta}
    + \frac{\gamma_\nu (\slashed{q}_1 - \slashed{k}_1)\gamma_\mu}
    {(q_1-k_1)^2 + i\eta} \biggr] v^{s_2}(q_2)\,.
  \end{split}
\end{equation}
In the center-of-the-mass frame or the back-to-back kinematics, we can
write the photon momenta, $\bk_1=-\bk_2=\bk$, and the fermion momenta,
$\bq_1=-\bq_2=\bq$.  Without loss of generality we can choose
$\hat{\bk}$ along the $z$-axis, $\bk\propto\be_z$, so that the
polarization vectors are 
$\bepsilon_\sigma=[\mathrm{sgn}(k_z)\be_x-i\sigma\be_y]/\sqrt{2}$. 
For massless particles, $\omega=|\bq|=|\bk|=|k_z|$ should hold, and we
introduce the scattering angle, i.e., $\bk\cdot\bq=\omega^2\cos\theta$ as well as
the azimuthal angle $\varphi$.
For the fermion polarization, in the chiral representation, we adopt
the standard choice of the left-handed and the right-handed fermions as
\begin{equation}
  u^L(q) = \sqrt{2\omega} \binom{\xi_{\downarrow}}{0}\,,\quad 
  u^R(q) = \sqrt{2\omega} \binom{0}{\xi_{\uparrow}}\,.
\end{equation}
For the negative energy states, $v=i\gamma^2 u^\ast$ gives
\begin{equation}
  v^{L}(q) = \sqrt{2\omega} \binom{0}{-\xi_{\uparrow}}\,,\quad  
  v^{R}(q) = \sqrt{2\omega} \binom{-\xi_{\downarrow}}{0}\,,  
\end{equation}  
where
$\xi_{\uparrow}=(\cos\frac{\theta}{2},
e^{i\varphi}\sin\frac{\theta}{2})^T$
and $\xi_{\downarrow}=(-e^{-i\varphi}\sin\frac{\theta}{2},
\cos\frac{\theta}{2})^T$.
Here, we must be careful about the momentum direction.  In our
convention $q_2$ of the argument $v^{s_2}(q_2)$ has $-\bq$ in the
spatial part, and this means that we should change the variables as
$\varphi\to\varphi+\pi$ and $\theta\to\pi-\theta$ for $v^{s_2}(q_2)$.
Therefore, using $q'=(\omega,-\bq)$, the anti-particle should take
\begin{equation}
  v^{L}(q') = \sqrt{2\omega} \binom{0}{-\tilde{\xi}_{\uparrow}}\,,\quad  
  v^{R}(q') = \sqrt{2\omega} \binom{-\tilde{\xi}_{\downarrow}}{0}
\end{equation} 
with
$\tilde{\xi}_{\uparrow}=(\sin\frac{\theta}{2},-e^{i\varphi}\cos\frac{\theta}{2})^T
\propto \xi_{\downarrow}$
and
$\tilde{\xi}_{\downarrow}=(e^{-i\varphi}\cos\frac{\theta}{2},\sin\frac{\theta}{2})^T
\propto \xi_{\uparrow}$.
We note that $v^{L/R}$ denotes the left-handed and the right-handed
anti-particle.

After tedious but straightforward calculations, we find that only the
following matrix elements,
\begin{equation}
  \begin{split}
    & i\calM_{+1,-1}^{RL} = [i\calM_{-1,+1}^{LR}]^\ast =
    -2ie^2\, e^{-2i\phi} \tan\frac{\theta}{2}\,,\\
    & i\calM_{-1,+1}^{RL} = [i\calM_{+1,-1}^{LR}]^\ast =
    2ie^2\, e^{2i\phi} \cot\frac{\theta}{2}\,,
  \end{split}
\end{equation}
are nonzero and others are all zero, i.e.,
$i\calM_{+1,+1}^{RL}=i\calM_{-1,-1}^{RL}=i\calM_{+1,+1}^{LR}=i\calM_{-1,-1}^{LR}=0$ and
$i\calM_{\sigma_1 \sigma_2}^{LL}=i\calM_{\sigma_1 \sigma_2}^{RR}=0$.
As expected from the no-go theorem derived from the helicity
conservation, these channels with nonzero amplitude have vanishing net
helicity, so as to satisfy the helicity constraint.

\section{Magnetic Polarizer}
\label{sec:polarizer}

\begin{figure}
  \includegraphics[width=0.4\columnwidth]{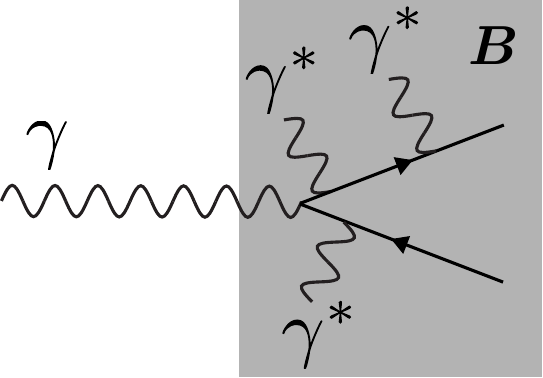}
  \caption{Schematic figure of scattering processes of 
    $\gamma\ell+\bar{\ell}$ under a uniform background field $\bB$.}
  \label{fig:B}
\end{figure}

So far, we have argued that the chirality production is prohibited in
on-shell photon annihilation process in which the quantized magnetic
helicity cannot be saturated by the matter chirality.  However, as
argued in the introduction, it is a well-known mechanism that the
chirality can be photo-induced by the presence of background fields of
parallel $\bE$ and $\bB$.  This is possible because the magnetic
helicity is not quantized for classical background field.  In other
words the multiple scattering of off-shell photons can yield
non-integer magnetic helicity.

We already mentioned in the introduction that the Schwinger mechanism
is essentially a nonperturbative tunneling phenomenon and it is still
difficult to achieve such a strong $\bE$ above the pair production
threshold experimentally.  Then, an intriguing and alternative
possibility is a situation in which we can replace $\bE$ with an
energetic on-shell photon while keeping $\bB$ as an external field.

\subsection{Considerations from the helicity conservation}

\begin{figure}
  \includegraphics[width=0.7\columnwidth]{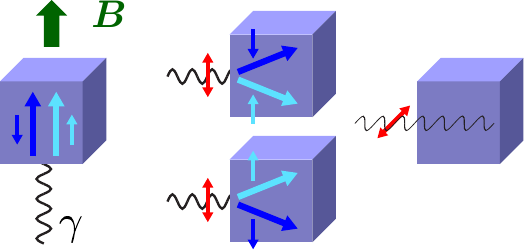}
  \caption{Configurations of the parallel (left) and the
    orthogonal (right) injection of $\gamma$ to a medium under strong $\bB$.  The orthogonally injected $\gamma$ is classified into two linear polarization.}
  \label{fig:translong}
\end{figure}

Now, let us focus on the simple process of
$\gamma \to \ell(B) + \bar{\ell}(B)$ where $\ell(B)$ and $\bar{\ell}(B)$
represent a fermion and an anti-fermion in the magnetic field.  We further specify the geometrical configuration in the
following way;  $\bB$ is imposed along the positive $z$ direction and
the on-shell photon, $\gamma$, is injected along either the $z$ axis (that is parallel to $\bB$ as shown in the left panel of
Fig.~\ref{fig:translong}) or the $x$ axis (that is orthogonal to $\bB$ as shown in the
right panel of Fig.~\ref{fig:translong}).  Furthermore, we shall
simplify the analysis assuming that $|\bB|$ is sufficiently strong to
justify the lowest Landau level (LLL) approximation.

For the parallel injection, we can understand no chirality
production from an intuitive argument; that is, the momentum
conservation along the magnetic direction should hold even in the
presence of $\bB$, which leads to $k^z=q_1^z+q_2^z > 0$.  At the same
time the energy conservation should be imposed since no work is
transferred from $\bB$ to charged particles.  Therefore, the kinematic
constraint is $|k^z|=\varepsilon_{n_1,q_1^z}+\varepsilon_{n_2,q_2^z}$,
where $\varepsilon_{n,q}=\sqrt{2eB n + q^2}$ represents the Landau quantized
energy.  They can be simultaneously satisfied only when two fermions
belong to the LLL (and the LLL approximation is not required here).
For negatively charged fermions like electrons, the particle in the
LLL has only the negative helicity (or the $L$ fermion) and the
anti-particle like positrons in the LLL has only the positive helicity
(or the anti-fermion of $L$, that is the $R$ anti-fermion) as depicted by the bigger arrows (momentum)
and the smaller arrows (spin) for particle (dark blue) and
anti-particle (light blue) in the left panel of
Fig.~\ref{fig:translong}.  Therefore, the net chirality should be
zero, and for this configuration the helicity conservation cannot
impose any interesting constraint.

The orthogonal injection is much more interesting since this
configuration allows for nonzero chirality production, which can also
be understood intuitively.  As shown in the middle panel of
Fig.~\ref{fig:translong}, the momentum conservation in the $z$
direction results in a pair with $q_1^z = -q_2^z$.  The spin alignment
along $\bB$ is preferred and the full polarization is realized in the
LLL approximation.  For $q_1^z > 0$ (that is, a produced particle
moves in the positive $z$ direction) a pair of the $L$ (negatively
charged) fermion and the $L$ anti-fermion can be produced with
$\Delta N_5=-2$.  At the same time, we should take account of another possible
process for $q_1^z<0$ and the production of a pair of the $R$ fermion
and the $R$ anti-fermion with $\Delta N_5=+2$ occurs at the same
rate.  In total, therefore, net chirality production is vanishing in
the average over time.  Here, the important point is that,
in the LLL approximation under strong enough $\bB$, the pair
production inevitably involves processes with $\Delta N_5=\pm 2$, and
this should be permitted microscopically by the helicity
conservation.  This condition leads to a conclusion that the incident
photon must be linear polarized along the $z$ direction that is
parallel to $\bB$.  In other words, if the incident photon is linear
polarized in a direction perpendicular to the $z$ direction (i.e., the
$y$ direction) as sketched in the far right panel of
Fig.~\ref{fig:translong}, the pair production is prohibited in the LLL
approximation.  Accordingly, the magnetic field turns out to be a polarizer!  We
note that Ref.~\cite{Hirono:2012ki} addressed a similar idea for the
polarizer with vortices at finite density (or finite chemical
potential $\mu$) and rotation (or angular velocity $\omega$).  Here,
we point out that $\bB$ by itself is already a polarizer, and this is
not astonishing if we recall that a correspondence, $B\sim \mu\omega$,
holds in various calculations as argued, for instance, in
Ref.~\cite{Stephanov:2012ki}.

Theoretically speaking, this polarization dependence of the pair
production is a consequence from the imaginary-part of the self-energy
in $\bB$, while the real-part results in the vacuum birefringence
induced by $\bB$ (see Refs.~\cite{Hattori:2012je,Hattori:2012ny} for
concrete calculations).  For the imaginary-part evaluation in the LLL
approximation, see also Ref.~\cite{Fukushima:2011nu}.  From the point
of view of physics applications, recently, the polarization dependent
opacity has been intensively discussed especially in the context of
the thermal X-ray from the magnetar.  In the plasma physics the mode
with $\bE\parallel \bB$ as in the middle panel in
Fig.~\ref{fig:translong} is called the ordinary mode (O-mode) and the
mode with $\bE\perp \bB$ as in the right panel is called the
extraordinary mode (X-mode), and it is known that the opacity of the
X-mode is smaller than that of the O-mode; see Ref.~\cite{Lai:2002ni}
for example for discussions in connection to the magnetar.  See also
Ref.~\cite{Lai:2022knd} for a possible interpretation for the
astrophysical measurement~\cite{doi:10.1126/science.add0080}.

In this sense, our observation of the magnetic field as a polarizer had been recognized in some contexts, but our emphasis in this work lies in the
implication of the chiral anomaly and the helicity conservation guaranteed by the axial Ward
identity.  At a first glance, the polarization
dependent opacity and the chiral anomaly seem to be completely distinct phenomena, and indeed they can be calculated out independently.  Nevertheless, once the helicity conservation is
established from the chiral anomaly, we can easily predict the selection rule for the
microscopic process of the particle production according to the
chirality carried by the particles.  In the end, it is instructive that the magnetic property as a
polarizer can naturally be understood from a simple principle of the helicity conservation
even without explicit calculation.

\subsection{Calculations in the LLL approximation}

Finally, we will present concrete calculations for the chirality
production amplitude in the LLL approximation.   For this purpose it
is convenient to introduce the spinors expressed in the cylindrical
coordinates $(\rho,\varphi,z)$ in the presence of $\bB$ along the $z$
axis.  Then, for the LLL, we can just consider the state with the
negative polarization (i.e., anti-parallel to $\bB$) with the $z$
component of the angular momentum, $j_z=l+\frac{1}{2}$, which is given
by~\cite{Fukushima:2020ncb}
\begin{equation}
  \begin{split}
    &u_{n,l,q^z}^{(\downarrow)}(\rho,\varphi,z,t)  
    = \frac{e^{-i\varepsilon_{n,l,q^z}^{(\downarrow)}t+iq^z z}}
    {\sqrt{\varepsilon_{n,l,q^z}^{(\downarrow)}}} \\
    &\times \begin{pmatrix}
      0\\
      \varepsilon_{n,l,q^z}^{(\downarrow)} \Phi_{n,l}(\chi^2,\varphi)\\
      -i\sqrt{|eB| (2n\!+\!|l\!+\!1|\!+\!l\!+\!1)} \Phi_{n,l+1}(\chi^2,\varphi)\\ 
      -q^z \Phi_{n,l+1}(\chi^2,\varphi)  
    \end{pmatrix}
  \end{split}
\end{equation}
in the Dirac representation of the $\gamma$ matrices.  Here, the
dimensionless variable is $\chi^2=\frac{1}{2}|eB| \rho^2$ and the special
function, $\Phi_{n,l}(\chi^2,\varphi)$, is defined by the associate Laguerre function, $L_n^m(x)$ as
\begin{equation}
  \Phi_{n,l}(\chi^2,\varphi) = \sqrt{\frac{n!}{(n+|l|)!}}
  e^{-\frac{1}{2}\chi^2} \chi^{|l|} L_n^{|l|}(\chi^2) e^{il\varphi}\,. 
\end{equation}
The Landau quantized energy dispersion relation is
$\varepsilon_{n,l,q}^{(\downarrow)} =\sqrt{|eB|(2n+|l+1|+l+1)+q^2}$.
From this expression we see that the LLLs exist for $n=0$ and
$l\le -1$ (or $j_z\le -\frac{1}{2}$).  Similarly, we can deal with
$u_{n,l,q^z}^{(\uparrow)}$ but these states are irrelevant for the
LLL approximation.  Within the LLL approximation, another interesting
state is the anti-fermion, $v_{n,l,q^z}^{(\uparrow)}$, given by
\begin{equation}
  \begin{split}
    &v_{n,l,q^z}^{(\uparrow)}(\rho,\varphi,z,t) 
    = \frac{e^{i\bar{\varepsilon}_{n,l,q^z}^{(\uparrow)}t-iq^z z}}
    {\sqrt{\bar{\varepsilon}_{n,l,q^z}^{(\uparrow)}}} \\
    &\times \begin{pmatrix}
      -i\sqrt{|eB| (2n\!+\!|l|\!-\!l)} \Phi_{n,-l-1}(\chi^2,\varphi)\\
      -q^z \Phi_{n,-l}(\chi^2,\varphi)\\
      0\\
      \bar{\varepsilon}_{n,l,q^z}^{(\uparrow)} \Phi_{n,-l}(\chi^2,\varphi)
    \end{pmatrix}
  \end{split}
\end{equation}
with
$\bar{\varepsilon}_{n,l,q}^{(\uparrow)}=\sqrt{|eB|(2n+|l|-l)+q^2}$.
This means that the LLLs are found for $n=0$ and $l\ge 0$ (or
$j_z \ge \frac{1}{2}$).

\begin{figure}
    \centering
    \includegraphics[width=\linewidth]{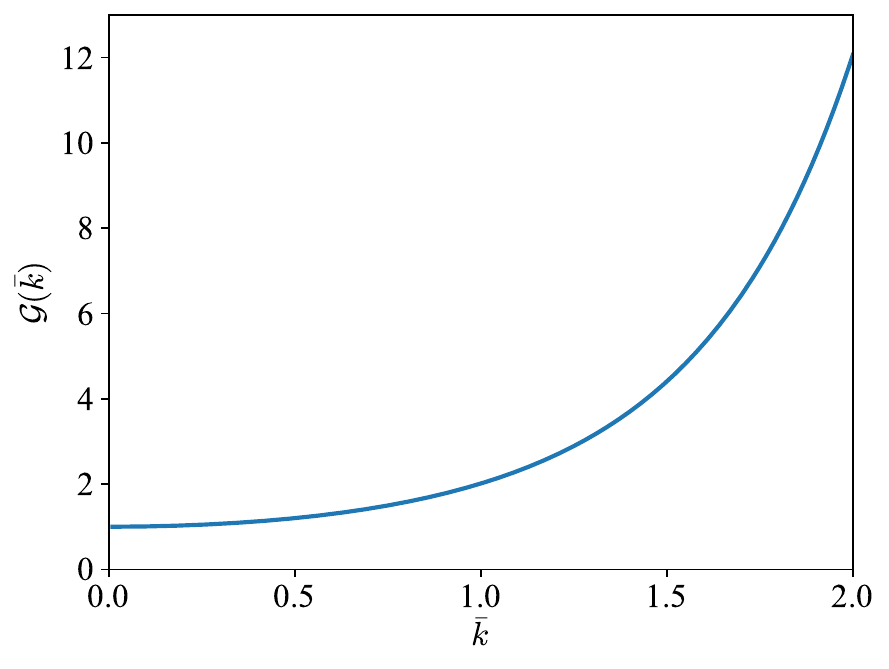}
    \caption{$\mathcal{G}(\bar{k})$ shown as a function of the rescaled photon momentum, $\bar{k}=k\sqrt{2/eB}$.}
    \label{fig:G}
\end{figure}

For this configuration, we choose the polarization vector of the
photon propagating in the $x$ direction as
$\bepsilon_+=(0,1,i)$.
Then, the amplitude for the chirality production in
the LLL approximation is
\begin{align}
  & -ie \frac{|eB|}{2\pi}\int d^4 x e^{-i(\omega t - kx)}\,
    \bar{u}_{0,l,q_1^z}^{(\downarrow)} (\gamma^2 + i\gamma^3)
    v_{0,l',q_2^z}^{(\uparrow)} \notag\\
  & = (2\pi)^{2}\delta(|q_{1}^{z}|+|q_{2}^{z}|-\omega)\delta(q_{1}^{z}+q_{2}^{z})i\mathcal{M}_{l,l'}(k)\,,
\end{align}
where $\omega=k=|\bk|$ is the energy of the photon.
The overall factor, $|eB|/(2\pi)$, appears from the normalization convention of the above spinors.
Hereafter, we simply write $eB$ to mean $|eB|$ for notation brevity.
We introduced the invariant amplitude as
\begin{equation}
    i\mathcal{M}_{l,l'}(\omega)
    = \frac{e^2 B\omega}{4\pi}\int \rho d\rho d\varphi\, e^{ik\rho\cos\varphi}\,
    (\Phi_{0,l+1} - \Phi_{0,l}) \Phi_{0,-l'}
\end{equation}
with $\Phi_{0,l}= e^{-\frac{1}{2}\chi^2}
\chi^{|l|} e^{il\varphi}/\sqrt{|l|!}$ and
$l\le -1$ and $l'\ge 0$.
From the explicit form of the spinors, we see that the above amplitude is proportional to $|q_1^z|=|q_2^z|$ and we replaced it with $\omega/2$ using the energy-momentum conservation.
To simplify the notation, we shall use a dimensionless variable, $\bar{k} = k\sqrt{2/eB}$, to find
\begin{equation}
    i\mathcal{M}_{l,l'}(\bar{k})=e\omega
    \bigl[
    \mathcal{F}_{l+1,l'}(\bar{k}) - \mathcal{F}_{l,l'}(\bar{k}) \bigr]\,,
\end{equation}
where
\begin{align}
    \mathcal{F}_{l,l'} &= \frac{(i)^{l-l'}}{\sqrt{|l|!|l'|!}}
    \int_0^\infty d\chi\,\chi^{-l+l'+1} e^{-\chi^2}
    J_{l-l'}(\bar{k}\chi) \notag\\
    &= \frac{(i \bar{k}/2)^{l'-l}}{2\sqrt{|l|! |l'|!}}
    e^{-\bar{k}^2/4} \,.
    \label{eq:F}
\end{align}
Let us consider the chirality production for the kinematics, $q_1^z=\omega/2$ and $q_2^z=-\omega/2$, that is, the fermion density from this process $\mathfrak{d}$ averaged over the transverse area $S_\perp$ is
\begin{align}
    \mathfrak{d} \Bigr|_{q_1^z > 0}
    &= \biggl(\frac{1}{S_\perp}\frac{2\pi}{eB}\biggr)
    \frac{1}{2\omega\times 2|q_1^z|\times 2|q_2^z|} \notag\\
    & \qquad \times \biggl(\frac{eB}{2\pi}\biggr)^2 \sum_{l\le-1, l'\ge 0}\bigl|i\mathcal{M}_{l,l'}(\bar{k})\bigr|^2 \notag\\
    &= \frac{eB}{2\pi} \frac{e^{-\omega^2/eB}}{8S_\perp \omega}
    \mathcal{G}(\bar{k})\,.
\end{align}
The function $\mathcal{G}(\bar{k})$ involves the summation with respect to $l$ and $l'$, which cannot be expressed in a closed form.  Thanks to $1/\sqrt{|l|!}$ in the normalization, the summation converges and we show the quantitative behavior in Fig.~\ref{fig:G}.
From the figure we see that $\mathcal{G}(\bar{k})$ is a monotonically increasing function with increasing $\bar{k}$.  Interestingly, the increase is not a simple power law, but the increasing rate grows for larger $\bar{k}$.  This is quite reasonable;  for large $k$ or $\omega$, the photon can interact with larger $l$, $l'$ modes, so that the power of $\bar{k}$ would become larger according to Eq.~\eqref{eq:F}.
It should be noted that the process with $q_1^z<0$ has the same amplitude, i.e., $\mathfrak{d} \Bigr|_{q_1^z<0} = \mathfrak{d} \Bigr|_{q_1^z>0}$.
Thus, each microscopic process allows for the positive or negative chirality production, and the net chirality averaged over time is unchanged. 

\section{Conclusions}
\label{sec:conclusions}

We discussed the implications of the helicity conservation that is imposed by the chiral anomaly through the axial Ward identity.
The helicity conservation describes the conversion between the chirality in the matter sector and the magnetic helicity in the gauge sector, as is seen from Eq.~\eqref{eq:axialWI}.
We also made a brief overview about the helicity conservation as known in optics, hoping that our consideration based on the quantum field theory may find a useful application in optics.

The helicity conservation consists of the chirality charge, $N_5$, and the magnetic helicity, $H_M$, apart from the spatial part.
We realized that in on-shell scattering processes both $N_5$ and $H_M$ are integers; $N_5,\, H_M \in \mathbb{Z}$.
They are connected with a non-integer coefficient, $e^2/(4\pi^2)$.
Therefore, the process consistent with the helicity conservation must satisfy $\Delta N_5 = \Delta H_M = 0$ unless $e^2/(4\pi^2)$ is a rational number and $H_M$ happens to make $(e^2/4\pi^2)\Delta H_M$  be integer.
In this way we reached a conclusion that the annihilation process of on-shell $\gamma$'s can produce pairs of fermions and anti-fermions but cannot generate non-zero chirality.

Our statement can be regarded as a no-go theorem.
To demonstrate this conclusion with explicit calculations, we considered the tree-level diagram for
$\gamma+\gamma\to\ell+\bar{\ell}$.
This is a textbook exercise if the polarization sum is taken, but we took one step further to compute the scattering amplitude for all the combinations of the incoming photon helicity and the outgoing fermion spin.
We then confirmed that the scattering amplitude is certainly nonzero only in the channels with $\Delta N_5=\Delta H_M=0$.

Finally, we considered a similar but different setup of
$\gamma \to \ell(B) + \bar{\ell}(B)$,
where $\ell(B)$ and $\bar{\ell}(B)$ represent the fermion and the anti-fermion with multiple scatterings with off-shell photons from the external magnetic field.
In this case the medium opacity in response to incident $\gamma$ depends on the relative direction between $\gamma$ and $\bB$ as well as the angle of the linear polarization.
In particular, when $\gamma$ is injected in the direction orthogonal to $\bB$, the opacity is larger if the linear polarization of $\gamma$ is chosen along the $\bB$ direction.
We can understand this tendency from the helicity conservation;  if $\bB$ is strong enough to justify the LLL approximation, the fluctuation of nonzero $\bE\cdot\bB$ (where $\bE$ originates from polarized $\gamma$) can be converted to the pair produced particles.
In this sense we can say that the magnetic medium plays a role as a polarizer, which we call the magnetic polarizer.
It would be an interesting test to conduct table-top experiments to measure the polarization efficiency using Weyl- and Dirac-semimetals in the magnetic field.

Prospective future extensions of the present work include:
1) The calculation of corrections due to fermion mass, which enables us to approach the situation relevant to analyses in optics.
2) The reformulation of Eq.~\eqref{eq:opt_cont} based on the quantum field theory in a way similar to the axial Ward identity (with both $\bA$ and $\bC$).
3) Refined evaluation of the particle production with $\gamma$ injection under $\bB$ beyond the LLL approximation, which would be crucial for comparison with experimental data.
4) Interplay between the helicity conservation and the angular momentum conservation for which not only the spin (helicity) but also the orbital angular momentum should be taken into account.
5) Related to this interplay, it would be challenging to perform the theoretical studies of the scattering involving paraxial photons, that is, the photon vortex beam with nonzero orbital angular momentum.


\begin{acknowledgments}
The authors thank
Yoshimasa Hidaka for useful conversations including paraxial photons,
and Takuya~Shimazaki,
for discussions about mode decomposed description.
This work was supported by Japan Society for the Promotion of Science
(JSPS) KAKENHI Grant Nos.\ 22H01216 (K.F.) and 22H05118 (K.F.).
\end{acknowledgments}        

\bibliography{paraxial}
\bibliographystyle{apsrev4-2}
\end{document}